\documentclass[10pt,a4paper,english,conference]{IEEEtran}
\usepackage[T1]{fontenc}
\usepackage[latin9]{inputenc}
\usepackage{amsthm}
\usepackage{amsmath}
\usepackage{amssymb}
\usepackage{graphicx}

\makeatletter

  \theoremstyle{plain}
  \newtheorem{thm}{\protect\theoremname}
  \theoremstyle{remark}
  \newtheorem{rem}{\protect\remarkname}
 \ifx\proof\undefined\
   \newenvironment{proof}[1][\proofname]{\par
     \normalfont\topsep6\p@\@plus6\p@\relax
     \trivlist
     \itemindent\parindent
     \item[\hskip\labelsep
           \scshape
       #1]\ignorespaces
   }{%
     \endtrivlist\@endpefalse
   }
   \providecommand{\proofname}{Proof}
 \fi
  \theoremstyle{definition}
  \newtheorem{example}{\protect\examplename}

\makeatother

\usepackage{babel}
\providecommand{\examplename}{Example}
\providecommand{\remarkname}{Remark}
\providecommand{\theoremname}{Theorem}

\begin{document}
\IEEEoverridecommandlockouts

\title{On the Role of Common Codewords in Quadratic Gaussian Multiple Descriptions 
Coding}

\author{Kumar Viswanatha, Emrah Akyol and Kenneth Rose\\
ECE Department, University of California - Santa Barbara\\
\{kumar,eakyol,rose\}@ece.ucsb.edu\thanks{The work was supported by the NSF under grants CCF - 1016861 and  CCF-1118075.} }
\maketitle
\begin{abstract}
This paper focuses on the problem of $L-$channel quadratic Gaussian
multiple description (MD) coding. We recently introduced a new encoding
scheme in \cite{our_ISIT} for general $L-$channel MD problem, based
on a technique called `Combinatorial Message Sharing' (CMS), where
every subset of the descriptions shares a distinct common message.
The new achievable region subsumes the most well known region for
the general problem, due to Venkataramani, Kramer and Goyal (VKG)
\cite{VKG}. Moreover, we showed in \cite{our_ITW} that the new scheme
provides a strict improvement of the achievable region for any source
and distortion measures for which some 2-description subset is such
that the Zhang and Berger (ZB) scheme achieves points outside the
El-Gamal and Cover (EC) region. In this paper, we show a more surprising
result: CMS outperforms VKG for a general class of sources and distortion
measures, which includes scenarios where for all 2-description subsets,
the ZB and EC regions coincide. In particular, we show that CMS strictly
extends VKG region, for the $L$-channel quadratic Gaussian MD problem
for all $L\geq3$, despite the fact that the EC region is complete
for the corresponding 2-descriptions problem. Using the encoding principles
derived, we show that the CMS scheme achieves the complete rate-distortion
region for several asymmetric cross-sections of the $L-$channel quadratic
Gaussian MD problem, which have not been considered earlier. \end{abstract}

\section{Introduction\label{sec:Introduction}}

The multiple descriptions (MD) problem has been studied extensively,
yielding a series of advances , ranging from achievability \cite{EGC,ZB,VKG,Ramchandran,our_ISIT,our_ITW,Binned_CMS_ITW}
to converse results \cite{Ozarow,Jun_Chen_ind_central,Viswanatha_2_levels}.
In the general MD setup, the encoder generates $L$-descriptions of
the source for transmission over $L$ available channels and it is
assumed that the decoder receives a subset of the descriptions perfectly
and the remaining are lost. The objective is to quantify the set of
all achievable rate-distortion (RD) tuples for the $L-$rates $(R_{1},\ldots,R_{L})$
and distortion levels corresponding to the $2^{L}-1$ possible description
loss patterns $(D_{\mathcal{K}},\mathcal{K}\subseteq\{1,\ldots,L\})$.
One of the first achievable regions for the 2-channel MD problem was
derived by El-Gamal and Cover (EC) in 1982 \cite{EGC}. It was shown
by Ozarow in \cite{Ozarow} that the EC region is complete when the
source is Gaussian and the distortion measure is mean squared error
(MSE). Zhang and Berger (ZB), however, later showed in \cite{ZB}
that the EC coding scheme is strictly sub-optimal in general. In particular,
for a binary source under Hamming distortion, sending a common codeword
within the two descriptions can achieve points that are strictly outside
the the EC region. The converse to the ZB scheme in still not known
for general sources and distortion measures.

Since then several researchers have worked on extending the EC and
ZB approaches to the $L-$channel MD problem \cite{VKG,Ramchandran,Jun_Chen_ind_central,Viswanatha_2_levels}.
An achievable scheme, due to Venkataramani, Kramer and Goyal (VKG)
\cite{VKG}, directly builds on EC and ZB, and introduces a combinatorial
number of refinement codebooks, one for each subset of the descriptions.
Motivated by ZB, a \textit{single} common codeword is also shared
by all the descriptions, which assists in better coordination of the
messages, improving the RD trade-off. We recently introduced a new
coding scheme called `Combinatorial Message Sharing' (CMS) in \cite{our_ISIT},
wherein a distinct common codeword is shared by members of each subset
of the transmitted descriptions. The new achievable RD region subsumes
the VKG region for general sources and distortion measures. Moreover,
we demonstrated in \cite{our_ITW} that CMS achieves a strictly larger
region than VKG for all $L>2$, if there exists a 2-description subset
for which ZB achieves points strictly outside the EC region. In particular,
CMS achieves strict improvement for a binary source under Hamming
distortion. 

Ozarow's converse result \cite{Ozarow} motivated researchers to seek
extended results for the $L-$channel quadratic Gaussian MD problem
\cite{Jun_Chen_ind_central,Viswanatha_2_levels}. It was shown in
\cite{Jun_Chen_ind_central} that a special case of the VKG coding
scheme, called the `correlated quantization' scheme (a generalization
of Ozarow's encoding mechanism to $L-$channels), where \textit{no
common codewords are sent,} achieves the complete rate region, when
only the individual and the central distortion constraints are imposed.
A different and important line of attack focused on a practically
interesting cross-section of the general MD problem, called the `symmetric
MD problem' (see \cite{Ramchandran}), based on encoding principles
derived from Slepian and Wolf's random binning techniques. In fact,
CMS principles can be extended to incorporate such random binning
techniques, to utilize the underlying symmetry in the problem setup
as illustrated recently in \cite{Binned_CMS_ITW}. However, in this
paper, we restrict ourselves to the general asymmetric setup to demonstrate
the potential gains of using the common codewords of CMS for the quadratic
Gaussian MD problem. 

Optimality of EC for the 2-descriptions setup has led to a natural
conjecture that common codewords do not play a necessary role in quadratic
Gaussian MD coding, and all the achievable regions characterized so
far neglect the common layer codewords (see eg., \cite{VKG,Jun_Chen_ind_central,Viswanatha_2_levels}).
In this paper, we show that, surprisingly CMS strictly outperforms
VKG for a Gaussian source under MSE distortion. More generally, we
show that strict improvement holds for a general class of sources
and distortion measures, which includes several scenarios in which,
for every 2-description subset, ZB and EC lead to the same achievable
region. We also show that the common codewords of CMS play a critical
role in achieving the complete RD region for several asymmetric cross-sections
of the $L-$channel quadratic Gaussian MD problem. 

We note that, due to severe space constraints, in this paper, we avoid
restating all the prior results and refer the readers to \cite{our_ISIT}
and \cite{our_ITW} for a brief description of EC, ZB and VKG schemes.
In the following section, we begin with a brief description of the
CMS coding scheme.

\section{Formal definition and CMS coding scheme\label{sec:Prior_results-1}}

A source produces a sequence of $n$ iid random variables, denoted
by $X^{n}=\left(X^{(1)},X^{(2)}\ldots,X^{(n)}\right)$. We denote
$\mathcal{L}=\{1,\ldots,L\}$. There are $L$ encoding functions,
$f_{l}(\cdot)\,\, l\in\mathcal{L}$, which map $X^{n}$ to the descriptions
$J_{l}=f_{l}(X^{n})$, where $J_{l}\in\{1,\ldots B_{l}\}$ for some
$B_{l}>0$. The rate of description $l$ is defined as $R_{l}=\log_{2}(B_{l})$.
Each of the descriptions are sent over a separate channel and are
either received at the decoder error free or are completely lost.
There are $2^{L}-1$ decoding functions for each possible received
combination of the descriptions $\hat{X}_{\mathcal{K}}^{n}=\left(\hat{X}_{\mathcal{K}}^{(1)},\ldots,\hat{X}_{\mathcal{K}}^{(n)}\right)=g_{\mathcal{K}}(J_{l}:l\in\mathcal{K})$,
$\forall\mathcal{K}\subseteq\mathcal{L},\mathcal{K}\neq\phi$, where
$\hat{X}_{\mathcal{K}}$ takes on values on a finite set $\hat{\mathcal{X}}_{\mathcal{K}}$,
and $\phi$ denotes the null set. When a subset $\mathcal{K}$ of
the descriptions are received at the decoder, the distortion is measured
as $D_{\mathcal{K}}=E\left[\frac{1}{N}\sum_{t=1}^{n}d_{\mathcal{K}}(X^{(t)},\hat{X}_{\mathcal{K}}^{(t)})\right]$
for some bounded distortion measures $d_{\mathcal{K}}(\cdot)$ defined
as $d_{\mathcal{K}}:\mathcal{X}\times\hat{\mathcal{X}}_{\mathcal{K}}\rightarrow\mathcal{R}$.
A RD tuple $(R_{i},D_{\mathcal{K}}:i\in\mathcal{L},\mathcal{K}\subseteq\mathcal{L},\mathcal{K}\neq\phi)$
is achievable if there exit $L$ encoding functions with rates $(R_{1}\ldots,R_{L})$
and $2^{L}-1$ decoding functions yielding distortions $D_{\mathcal{K}}$.
The closure of the set of all achievable RD tuples is defined as the
`\textit{$L$-channel multiple descriptions RD region}'. 

In what follows, $2^{\mathcal{S}}$ denotes the set of all subsets
(power set) of any set $\mathcal{S}$ and $|\mathcal{S}|$ denotes
the set cardinality. Note that $|2^{\mathcal{S}}|=2^{|\mathcal{S}|}$.
$\mathcal{S}^{c}$ denotes the set complement. For two sets $\mathcal{S}_{1}$
and $\mathcal{S}_{2}$, we denote the set difference by $\mathcal{S}_{1}-\mathcal{S}_{2}=\{\mathcal{K}:\mathcal{K}\in\mathcal{S}_{1},\mathcal{K}\notin\mathcal{S}_{2}\}$.
We use the shorthand $\{U\}_{\mathcal{S}}$ for $\{U_{\mathcal{K}}:\mathcal{K}\in\mathcal{S}\}$%
\footnote{Note the difference between $\{U\}_{\mathcal{S}}$ and $U_{\mathcal{S}}$.
$\{U\}_{\mathcal{S}}$ is a set of variables, whereas $U_{\mathcal{S}}$
is a single variable. %
}. Before describing CMS, we define the following subsets of $2^{\mathcal{L}}$:
\begin{eqnarray}
\mathcal{I}_{W} & = & \{\mathcal{S}:\mathcal{S}\in2^{\mathcal{L}},\,\,|\mathcal{S}|=W\}\nonumber \\
\mathcal{I}_{W+} & = & \{\mathcal{S}:\mathcal{S}\in2^{\mathcal{L}},\,\,|\mathcal{S}|>W\}\label{eq:Iw}
\end{eqnarray}
Let $\mathcal{B}$ be any non-empty subset of $\mathcal{L}$ with
$|\mathcal{B}|\leq W$. We define the following subsets of $\mathcal{I}_{W}$
and $\mathcal{I}_{W+}$:
\begin{eqnarray}
\mathcal{I}_{W}(\mathcal{B}) & = & \{\mathcal{S}:\mathcal{S}\in\mathcal{I}_{W},\,\,\mathcal{B}\subseteq\mathcal{S}\}\nonumber \\
\mathcal{I}_{W+}(\mathcal{B}) & = & \{\mathcal{S}:\mathcal{S}\in\mathcal{I}_{W+},\,\,\mathcal{B}\subseteq\mathcal{S}\}\label{eq:Iwb}
\end{eqnarray}
We also define $\mathcal{J}(\mathcal{K})=\bigcup_{l\in\mathcal{K}}\mathcal{I}_{1+}(l)$.
Note that $\mathcal{J}(\mathcal{L})=2^{\mathcal{L}}-\phi$. 

Next, we briefly describe the CMS encoding scheme in \cite{our_ISIT}.
Recall that, unlike VKG, CMS allows for `combinatorial message sharing',
i.e a common codeword is sent in each (non-empty) subset of the descriptions.
The shared random variables are denoted by `$V$'. The base and the
refinement layer random variables are denoted by `$U$'. First, the
codebook for $V_{\mathcal{L}}$ is generated. Then, the codebooks
for $V_{\mathcal{S}}$, $|S|=W$ are generated in the order $W=L-1,L-2\ldots2$.
$2^{nR_{\mathcal{Q}}^{''}}$ codewords of $V_{\mathcal{Q}}$ are independently
generated conditioned on each codeword tuple of $\{V\}_{\mathcal{I}_{W+}(\mathcal{Q})}$.
This is followed by the generation of the base layer codebooks, i.e.
$U_{l}$, $l\in\mathcal{L}$. Conditioned on each codeword tuple of
$\{V\}_{I_{1+}(l)}$, $2^{nR_{l}^{'}}$ codewords of $U_{l}$ are
generated independently. Then the codebooks for the refinement layers
are formed by generating a single codeword for $U_{\mathcal{S}},\,\,|\mathcal{S}|>1$
conditioned on every codeword tuple of $(\{V\}_{\mathcal{J}(\mathcal{S})},\{U\}_{2^{\mathcal{S}}-\mathcal{S}})$.
Observe that the base and the refinement layers in the CMS scheme
are similar to that in the VKG scheme, except that they are now generated
conditioned on a subset of the shared codewords. 

The encoder employs joint typicality encoding, i.e., on observing
a typical sequence $x^{n}$, it tries to find a jointly typical codeword
tuple, one from each codebook. As with VKG, the codeword index of
$U_{l}$ (at rate $R_{l}^{'}$) is sent in description $l$. However,
now the codeword index of $V_{\mathcal{S}}$ (at rate $R_{\mathcal{S}}^{''}$)
is sent in \textit{all} the descriptions $l\in\mathcal{S}$. Therefore
the rate of description $l$ is:
\begin{equation}
R_{l}=R_{l}^{'}+\sum_{\mathcal{K}\in\mathcal{J}(l)}R_{\mathcal{K}}^{''}\label{eq:main_rate-1}
\end{equation}

We next formally state the achievable RD region. Let $\mathcal{Q}$
be any subset of $2^{\mathcal{L}}$. Then, we say that $\mathcal{Q}\in\mathcal{Q}^{*}$
if it satisfies the following property:
\begin{equation}
\mathcal{K}\in\mathcal{Q}\,\,\Rightarrow\,\,\,\mathcal{I}_{|\mathcal{K}|+}(\mathcal{K})\subset\mathcal{Q}\label{eq:cond_Q_main}
\end{equation}
$\forall\mathcal{K\in\mathcal{Q}}$. Further, we denote by $[\mathcal{Q}]_{1}$
the set of all elements of $\mathcal{Q}$ of cardinality $1$, i.e.,:
\begin{equation}
[\mathcal{Q}]_{1}=\{\mathcal{K}:\mathcal{K}\in Q,\,|\mathcal{K}|=1\}
\end{equation}

Let $(\{V\}_{\mathcal{J}(\mathcal{L})},\{U\}_{2^{\mathcal{L}}-\phi})$
be any set of $2^{L+1}-L-2$ random variables jointly distributed
with $X$. For any set $\mathcal{Q}\in\mathcal{Q}^{*}$ we define:
\begin{eqnarray}
\alpha(\mathcal{Q}) & = & \sum_{\mathcal{K}\in\mathcal{Q}-[\mathcal{Q}]_{1}}H\left(V_{\mathcal{K}}|\{V\}_{\mathcal{I}_{|\mathcal{K}|+}(\mathcal{K})}\right)\nonumber \\
 &  & +\sum_{\mathcal{K}\in2^{[\mathcal{Q}]_{1}}-\phi}H\left(U_{\mathcal{K}}|\{V\}_{\mathcal{I}_{1+}(\mathcal{K})},\{U\}_{2^{\mathcal{K}}-\phi-\mathcal{K}}\right)\nonumber \\
 &  & -H\left(\{V\}_{\mathcal{Q}-[\mathcal{Q}]_{1}},\{U\}_{2^{[\mathcal{Q}]_{1}}-\phi}|X\right)\label{eq:alpha_defn_1-1}
\end{eqnarray}
We follow the convention $\alpha(\phi)=0$. Next we state the rate-distortion
region achievable by the CMS scheme for the $L-$descriptions framework. 
\begin{thm}
\label{thm:main}Let $(\{V\}_{\mathcal{J}(\mathcal{L})},\{U\}_{2^{\mathcal{L}}-\phi})$
be any set of $2^{L+1}-L-2$ random variables jointly distributed
with $X$, where $U_{\mathcal{S}}$ and $V_{\mathcal{S}}$ take values
in some finite alphabets $\mathcal{U}_{\mathcal{S}}$ and $\mathcal{V}_{\mathcal{S}}$,
respectively $\forall\mathcal{S}$. Let $\mathcal{Q}^{*}$ be the
set of all subsets of $2^{\mathcal{L}}-\phi$ satisfying (\ref{eq:cond_Q_main})
and let $R_{\mathcal{S}}^{''},\,\,\mathcal{S}\in\mathcal{I}_{1+}$
and $R_{l}^{'},\,\, l\in\mathcal{L}$ be $2^{L}-1$ auxiliary rates
satisfying:

\textup{
\begin{equation}
\sum_{\mathcal{S}\in\mathcal{Q}-[\mathcal{Q}]_{1}}R_{\mathcal{S}}^{''}+\sum_{l\in[\mathcal{Q}]_{1}}R_{l}^{'}>\alpha(\mathcal{Q})\,\,\,\,\forall\mathcal{Q}\in\mathcal{Q}^{*}\label{eq:aux_rate_cond_thm}
\end{equation}
}Then, the RD region for the $L-$channel MD problem contains the
rates and distortions for which there exist functions $\psi_{\mathcal{S}}(\cdot)$,
such that, 
\begin{eqnarray}
R_{l} & = & R_{l}^{'}+\sum_{\mathcal{S}\in\mathcal{J}(l)}R_{\mathcal{S}}^{''}\label{eq:rate_condition_thm}\\
D_{\mathcal{S}} & \geq & E\left[d_{\mathcal{S}}\left(X,\psi_{\mathcal{S}}\left(U_{\mathcal{S}}\right)\right)\right]\label{eq:dist_condition_thm}
\end{eqnarray}
The closure of the achievable tuples over all such $2^{L+1}-L-2$
random variables is denoted by $\mathcal{RD}_{CMS}$.\end{thm}
\begin{rem}
$\mathcal{RD}_{CMS}$ can be extended to continuous random variables
and well-defined distortion measures using techniques similar to \cite{Wyner}.
We omit the details here and assume that the above region continues
to hold even for well behaved continuous random variables (for example,
a Gaussian source under MSE). 
\end{rem}

\begin{rem}
$\mathcal{RD}_{CMS}$ is convex, as a time sharing random variable
can be embedded in $V_{\mathcal{L}}$.\end{rem}
\begin{proof}
Refer to \cite{our_ISIT}.
\end{proof}

\section{Strict Improvement for a General Class of Sources and Distortion
Measures\label{sec:Prior_results}}

We begin by defining $\mathcal{Z}_{ZB}$, the set of all sources (for
given distortion measures at the decoders), for which there exists
an operating point $(R_{1},R_{2},D_{1},D_{2},D_{12})$ that \textit{cannot}
be achieved by an `independent quantization' mechanism using the ZB
coding scheme. More specifically, $X\in\mathcal{Z}_{ZB}$, if there
exists a strict suboptimality in the ZB region when the closure is
defined only over joint densities for the auxiliary random variables
satisfying the following conditions:
\begin{eqnarray}
P(U_{1},U_{2}|X,V_{12}) & = & P(U_{1}|X,V_{12})P(U_{2}|X,V_{12})\nonumber \\
E\left[d_{\mathcal{K}}(X,\psi_{\mathcal{K}}(U_{\mathcal{K}}))\right] & \leq & D_{\mathcal{K}},\,\;\;\;\mathcal{K}\in\{1,2,12\}\nonumber \\
U_{12} & = & f(U_{1},U_{2},V_{12})\label{eq:ZZB}
\end{eqnarray}
where $f$ is any deterministic function. We will show in Theorem
\ref{thm:General_CMS} that $\forall X\in\mathcal{Z}_{ZB}$, $\mathcal{RD}_{VKG}\subset\mathcal{RD}_{CMS}$. 

\begin{figure}
\centering\includegraphics[scale=0.4]{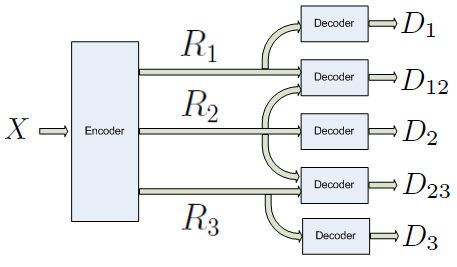}

\caption{The cross-section that we consider in order to prove that CMS achieves
points outside the VKG region for a general class of source and distortion
measures. CMS achieves the the complete RD region for this setup for
several distortion regimes for the quadratic Gaussian MD problem.
\label{fig:3_des_new}}
\end{figure}

Before stating the result we describe the particular cross-section
of the RD region that we will use to prove strict improvement in Theorem
\ref{thm:General_CMS}. Consider a 3-descriptions MD setup for a source
$X$ wherein we impose constraints only on distortions $(D_{1},D_{2},D_{3},D_{12},D_{23})$
and set the rest of the distortions, $(D_{13},D_{123})$ to $\infty$.
This cross-section is schematically shown in Fig. \ref{fig:3_des_new}.
To illustrate the gains underlying CMS, here we restrict ourselves
to the setting wherein we further impose $D_{1}=D_{3}$ and $D_{12}=D_{23}$.
The points in this cross-section, achievable by VKG and CMS, are denoted
by $\overline{\mathcal{RD}}_{VKG}(X)$ and $\overline{\mathcal{RD}}_{CMS}(X)$,
respectively. We note that the symmetric setting is considered \textit{only}
for simplicity. The arguments can be easily extended to the asymmetric
framework. 

This particular symmetric cross-section of the 3-descriptions MD problem
is equivalent to the corresponding 2-descriptions problem, in the
sense that, one could use any coding scheme to generate bit-streams
for descriptions 1 and 2, respectively. Description 3 would then carry
a replica (exact copy) of the bits sent in description 1. Due to the
underlying symmetry in the problem setup, the distortion constraints
at all the decoders are satisfied. Hence an achievable region based
on the ZB coding scheme can be derived as follows. Let $(G_{12},F_{1},F_{2},F_{12})$
be any random variables jointly distributed with $X$ and taking values
over arbitrary finite alphabets. Then the following RD-region is achievable
for which there exist functions $(\psi_{1},\psi_{2},\psi_{12})$ such
that $R_{1}=R_{3}$, $D_{1}=D_{3}$, $D_{12}=D_{23}$ and: 
\begin{eqnarray}
 & R_{1}\geq I(X;F_{1},G_{12}),\,\, R_{2}\geq I(X;F_{2},G_{12})\nonumber \\
 & R_{1}+R_{2}\geq2I(X;G_{12})+H(F_{1}|G_{12})+H(F_{2}|G_{12})\nonumber \\
 & -H(F_{1},F_{2},F_{12}|X,G_{12})+H(F_{12}|F_{1},F_{2},G_{12})\nonumber \\
 & D_{\mathcal{K}}\geq E\left[d_{\mathcal{K}}(X,\psi_{\mathcal{K}}(F_{\mathcal{K}}))\right],\,\,\mathcal{K}\in\{1,2,12\}\label{eq:RD(X)}
\end{eqnarray}
The closure of achievable RD-tuples over all random variables $(G_{12},M_{1},M_{2},M_{12})$
is denoted by $\overline{\mathcal{RD}}(X)$. In the following theorem,
we will show that $\overline{\mathcal{RD}}(X)\subseteq\overline{\mathcal{RD}}_{CMS}(X)$.
We also show that the VKG coding scheme \textit{cannot} achieve the
above RD region, i.e., $\overline{\mathcal{RD}}_{VKG}(X)\subset\overline{\mathcal{RD}}(X)$,
if $X\in\mathcal{Z}_{ZB}$. We note that in Theorem \ref{thm:General_CMS},
we focus only on the 3-descriptions setting. However, the results
can be easily extended to the general $L-$descriptions scenario.
Also note that $\overline{\mathcal{RD}}_{CMS}(X)$ could be strictly
larger than $\overline{\mathcal{RD}}(X)$, in general. 
\begin{thm}
\label{thm:General_CMS}(i) For the setup shown in Fig. \ref{fig:3_des_new}
the CMS scheme achieves $\overline{\mathcal{RD}}(X)$, i.e., $\overline{\mathcal{RD}}(X)\subseteq\overline{\mathcal{RD}}_{CMS}(X)$.

(ii) If $X\in\mathcal{Z}_{ZB}$,\textup{ }then there exists points
in $\overline{\mathcal{RD}}(X)$ that \textbf{cannot} be achieved
by the VKG encoding scheme, i.e., \textup{$\overline{\mathcal{RD}}_{VKG}\subset\overline{\mathcal{RD}}(X)$}, \end{thm}
\begin{rem}
It directly follows from (i) and (ii) that $\mathcal{RD}_{VKG}\subset\mathcal{RD}_{CMS}$
for the $L-$channel MD problem $\forall L\geq3$, if $X\in\mathcal{Z}_{ZB}$.\textit{ }\end{rem}
\begin{proof}
We first provide an intuitive argument to justify the claim and then
follow it up with a formal argument. Due to the underlying symmetry
in the setup CMS introduces common layer random variables $V_{123}=G_{12}$
and $V_{13}=F_{1}$. It then sends the codeword of $V_{13}$ is both
descriptions 1 and 3 (i.e., $U_{1}=U_{3}=V_{13}$). Hence it is sufficient
for the encoder to generate enough codewords of $U_{2}=F_{2}$ (conditioned
on $V_{123}$) to maintain joint typicality with the codewords of
$V_{13}=F_{1}$. However VKG is forced to set the common layer random
variable $V_{13}$ to a constant. Thus, in this case, the encoder
needs to generate enough number of codewords of $U_{2}$ so as to
maintain joint typicality individually with the codewords of $U_{1}$
and $U_{3}$, which are now generated independently conditioned on
$V_{123}$, entailing some excess rate for $U_{2}$%
\footnote{It might be tempting to conclude that the suboptimality in VKG is
due to conditions for joint typicality of all the codewords, while
for this cross-section, joint typicality of codewords of $U_{1}$
and $U_{3}$ is unnecessary. However, it is possible to show that
common codewords provide strict improvement even when joint typicality
only within prescribed subsets is imposed. The details are omitted
here. %
}. 

Part (i) of the theorem is straightforward to prove. We set $V_{123}=G_{12}$,
$V_{13}=F_{1}$, $U_{2}=F_{2}$, $U_{12}=U_{23}=F_{12}$ and $U_{1}=U_{3}=V_{13}$
and the rest of the random variables to constants in the CMS achievable
region in \cite{our_ISIT}. This leads to the RD region in (\ref{eq:RD(X)}).

We next prove (ii). We consider one particular boundary point of (\ref{eq:RD(X)})
and show that this cannot be achieved by VKG. Let $D_{1}$, $D_{2}$
and $D_{12}$ be fixed. Consider the following quantity:
\begin{eqnarray}
R_{VKG}^{*}(D_{1},D_{2},D_{12})=\inf\,\,\Bigl\{ R_{2}:R_{1}=R_{3}=R_{X}(D_{1})\label{eq:Thm3_2}\\
(R_{1},R_{2},R_{3},D_{1},D_{2},D_{1},D_{12},D_{12})\in\overline{\mathcal{RD}}_{VKG}(X)\Bigr\}\nonumber 
\end{eqnarray}
Note that the corresponding quantity achievable using $\overline{\mathcal{RD}}_{CMS}(X)$
is given by the solution to the following optimization problem:
\begin{eqnarray}
R_{CMS}^{*}(D_{1},D_{2},D_{12})=\inf\,\,\Bigl\{ I(U_{2};X,U_{1}|V_{123})\nonumber \\
I(X;V_{123})+I(U_{12};X|V_{123},U_{1},U_{2})\Bigr\}\label{eq:Thm3_3}
\end{eqnarray}
where the infimum is over all joint densities $P(V_{123},U_{1},U_{2},U_{12}|X)$,
where $P(V_{123},U_{1}|X)$ is any joint density for which there exists
a function $\psi_{1}(\cdot)$ such that:
\begin{eqnarray}
I(X;V_{123},U_{1})=R(D_{1}), &  & E\left[d_{1}(X,\psi_{1}(U_{1}))\right]=D_{1}\label{eq:CMS_13_constraints-1}
\end{eqnarray}
i.e., $(V_{123},U_{1})$ leads to an RD-optimal reconstruction of
$X$ at $D_{1}$ and $P(U_{12},U_{2}|X,U_{1},V_{123})$ is any distribution
for which there exists function $\psi_{2}(\cdot)$ and $\psi_{12}(\cdot)$
satisfying the distortion constraints for $D_{2}$ and $D_{12}$,
respectively. We will show that $R_{VKG}^{*}>R_{CMS}^{*}$. We next
specialize and restate $\overline{\mathcal{RD}}_{VKG}(X)$ for the
considered cross-section. Let $(V_{123},U_{1},U_{2},U_{3},U_{12},U_{23})$
be any random variables jointly distributed with $X$ taking values
on arbitrary alphabets. Then, $\overline{\mathcal{RD}}_{VKG}$ contains
all rates and distortions for which there exist functions $\psi_{1}(\cdot),\psi_{2}(\cdot),\psi_{3}(\cdot),\psi_{12}(\cdot),\psi_{23}(\cdot)$,
such that:
\begin{eqnarray}
R_{i} & \geq & I(X;U_{i},V_{123}),\,\, i\in\{1,2,3\}\nonumber \\
R_{i}+R_{2} & \geq & 2I(X;V_{123})+I(U_{i};U_{2}|V_{123})\nonumber \\
 &  & +I(X;U_{i},U_{2},U_{i2}|V_{123}),\,\, i\in\{1,3\}\nonumber \\
R_{1}+R_{3} & \geq & 2I(X;V_{123})+H(U_{1}|V_{123})\nonumber \\
 &  & +H(U_{3}|V_{123})-H(U_{1},U_{3}|X,V_{123})\nonumber \\
R_{1}+R_{2}+R_{3} & \geq & 3I(X;V_{123})+\sum_{i=1}^{3}H(U_{i}|V_{123})\nonumber \\
 &  & +\sum_{\mathcal{K}\in\{12,23\}}H(U_{\mathcal{K}}|\{U\}_{\mathcal{K}},V_{123})\nonumber \\
 &  & -H(U_{1},U_{2},U_{3},U_{12},U_{23}|X,V_{123})\label{eq:RD_VKG_bar-1}
\end{eqnarray}
\begin{eqnarray}
E\left(d_{\mathcal{K}}(X,\psi_{\mathcal{K}}(U_{\mathcal{K}}))\right) & \leq & D_{\mathcal{K}},\,\,\mathcal{K}\in\{1,2,3,12,13\}\label{eq:RD_VKG_Dist-1}
\end{eqnarray}
where $R_{1}=R_{3}$, $D_{1}=D_{3}$ and $D_{12}=D_{23}$. Observe
that the random variables $U_{13}$ and $U_{123}$ have been set to
constants as we do not impose distortion constraints $D_{13}$ and
$D_{123}$, respectively. We can further restrict the joint density
$P(V_{123},U_{1},U_{2},U_{3},U_{12},U_{23}|X)$ to satisfy:
\begin{eqnarray}
 & P(U_{12},U_{23}|X,V_{123},U_{1},U_{2},U_{3})=\nonumber \\
 & P(U_{12}|X,V_{123},U_{1},U_{2})P(U_{23}|X,V_{123},U_{2},U_{3})\label{eq:U_1223_Mark}
\end{eqnarray}
without any loss of optimality. 

Next imposing $R_{1}=R_{3}=R_{X}(D_{1})$ in (\ref{eq:RD_VKG_bar-1}),
enforces the joint density $P(V_{123},U_{1},U_{3}|X)$ to satisfy
the following constraints:
\begin{eqnarray}
I(X;V_{123},U_{i}) & = & R(D_{1}),\,\, i\in\{1,3\}\nonumber \\
E\left[d_{i}(X,\psi_{i}(V_{123},U_{i}))\right] & = & D_{1},\,\, i\in\{1,3\}\label{eq:VKG_13_constraints}\\
P(U_{1},U_{3}|X,V_{123}) & = & P(U_{1}|X,V_{123})\times P(U_{3}|X,V_{123})\nonumber 
\end{eqnarray}
where the last condition is required to satisfy the constraint on
$R_{1}+R_{3}$ in (\ref{eq:RD_VKG_bar-1}). Therefore, using (\ref{eq:RD_VKG_bar-1})
and (\ref{eq:U_1223_Mark}) we have:
\begin{eqnarray}
R_{VKG}^{*}=\inf\,\,\Bigl\{ I(X;V_{123})+I(U_{2};U_{1},U_{3},X|V_{123})\nonumber \\
+I(X;U_{12}|U_{1},U_{2},V_{123})+I(X;U_{23}|U_{2},U_{3},V_{123})\Bigr\}\label{eq:R_s_VKG}
\end{eqnarray}
where the infimum is over all joint densities $P(V_{123},U_{1},U_{2},U_{3},U_{12},U_{23}|X)$
satisfying (\ref{eq:VKG_13_constraints}) for which there exist functions
$\psi_{2}(\cdot),\psi_{12}(\cdot),\psi_{23}(\cdot)$ satisfying the
distortion constraints in (\ref{eq:RD_VKG_Dist-1}). 

From (\ref{eq:R_s_VKG}) and (\ref{eq:Thm3_3}) it follows that $R_{VKG}^{*}$
is equal to $R_{CMS}^{*}$ if and only if the two quantities on the
RHS of (\ref{eq:R_s_VKG}) and (\ref{eq:Thm3_3}), respectively, are
equal. However for any joint density, we have $I(U_{2};U_{1},U_{3},X|V_{123})\geq I(U_{2};U_{1},X|V_{123})$
and $I(X;U_{23}|V_{123},U_{2},U_{3})\geq0$. Also note that the constraints
in (\ref{eq:CMS_13_constraints-1}) are a subset of the constraints
in (\ref{eq:VKG_13_constraints}). Hence for $R_{VKG}^{*}$ to be
equal to $R_{CMS}^{*}$, any joint density which achieves $R_{VKG}^{*}$
must satisfy the following conditions:

(i) The joint density of $(X,V_{123},U_{1},U_{2},U_{12})$ must be
the same as the corresponding joint density which achieves $R_{CMS}^{*}$
(in (\ref{eq:Thm3_3})).

(ii)$I(U_{2};U_{3}|V_{123},U_{1},X)=0$, $I(X;U_{23}|V_{123},U_{2},U_{3})=0$.\\
The constraint $I(X;U_{23}|V_{123},U_{2},U_{3})=0$ implies that $X$
and $U_{23}$ are independent given $V_{123}$, $U_{2}$ and $U_{3}$.
Equivalently this constraint implies that the reconstruction $\hat{X}_{23}$
can be written as a deterministic function of $V_{123}$, $U_{2}$
and $U_{3}$, i.e., for $R_{VKG}^{*}$ to be equal to $R_{CMS}^{*}$,
there must exist a function $\tilde{\psi}_{23}(V_{123},U_{2},U_{3})$
such that $E\left(d_{23}(X,\tilde{\psi}_{23}(V_{123},U_{2},U_{3}))\right)\leq D_{23}=D_{12}$.
On the other hand, the constraint $I(U_{2};U_{3}|V_{123},U_{1},X)=0$
implies that $H(U_{3}|V_{123},U_{1},X)=H(U_{3}|V_{123},U_{1},U_{2},X)$.
However, the joint density of $(X,V_{123},U_{1},U_{3})$ must satisfy
(\ref{eq:VKG_13_constraints}) for $R_{1}=R_{3}=R_{X}(D_{1})$ to
hold, i.e., $H(U_{3}|V_{123},U_{1},X)=H(U_{3}|V_{123},X)$. Hence
for $R_{VKG}^{*}$ to be equal to $R_{CMS}^{*}$, we require:
\begin{equation}
H(U_{3}|V_{123},X)=H(U_{3}|V_{123},U_{1},U_{2},X)\label{eq:contra_cond2}
\end{equation}
which implies that $U_{2}\leftrightarrow(X,V_{123})\leftrightarrow U_{3}$
must hold. Recall that the joint density $P(U_{3},V_{123}|X)$ is
RD-optimal at $D_{1}$ and the joint density $P(U_{2},V_{123}|X,U_{1})$
must be identical to the joint density which achieves $R_{CMS}^{*}$
(from condition (i)). Hence, it follows that, if $X\in\mathcal{Z}_{ZB}$,
there exists at least one RD tuple in $\overline{\mathcal{RD}}(X)$
that cannot be achieved if we constrain the joint density to simultaneously
satisfy both the conditions (i) and (ii), proving the theorem.
\end{proof}
\textbf{Discussion:} A direct consequence of the above theorem is
that, if $X\in\mathcal{Z}_{ZB}$, then the common layer codewords
of CMS provide strict improvement in the achievable region as compared
not using them, i.e., if $X\in\mathcal{Z}_{ZB}$, $\mathcal{RD}_{VKG}\Bigr|_{V_{\mathcal{L}}=\Phi}\subseteq\mathcal{RD}_{VKG}\subset\mathcal{RD}_{CMS}$,
where $\mathcal{RD}\Bigr|_{V_{\mathcal{L}}=\Phi}$ denotes the VKG
region when the common layer random variable (denoted by $V_{\mathcal{L}}$)
is set to a constant $\Phi$%
\footnote{Note that setting $V_{\mathcal{L}}$ to a constant in VKG is equivalent
to setting all the common layer random variables to constants in CMS. %
}. In fact, it is possible to show that, whenever $X\in\mathcal{Z}_{EC}$,
$\mathcal{RD}\Bigr|_{V_{\mathcal{L}}=\Phi}\subset\mathcal{RD}_{CMS}$,
where $\mathcal{Z}_{EC}$ is defined as the set of all sources for
which there exists an operating point (with respect to the given distortion
measures) that \textit{cannot} be achieved by an `independent quantization'
mechanism using the EC coding scheme, i.e., if there exists an operating
point that \textit{cannot} be achieved by EC using a joint density
for the auxiliary random variables satisfying:
\begin{eqnarray}
P(U_{1},U_{2}|X) & = & P(U_{1}|X)P(U_{2}|X)\nonumber \\
E\left[d_{\mathcal{K}}(X,\psi_{\mathcal{K}}(U_{\mathcal{K}}))\right] & \leq & D_{\mathcal{K}},\,\,\,\mathcal{K}\in\{1,2,12\}\nonumber \\
U_{12} & = & f(U_{1},U_{2})\label{eq:ZEC}
\end{eqnarray}
where $f$ is any deterministic function. Note that the set $\mathcal{Z}_{ZB}$
is a subset of $\mathcal{Z}_{EC}$. Also observe that if $X\notin\mathcal{Z}_{EC}$,
the concatenation of two independent optimal quantizers is optimal
in achieving a joint reconstruction. While this condition could be
satisfied for specific values of $D_{1},D_{2}$ and $D_{12}$, it
is seldom achieved \textit{for all} values of $(D_{1},D_{2},D_{12})$.
Though such sources are of some theoretical interest, the multiple
descriptions encoding for such sources is degenerate. Hence with some
trivial exceptions, it can be asserted that the common layer codewords
in CMS can be used to achieve a strictly larger region (compared to
not using any common codewords) for all sources and distortion measures,
$\forall L\geq3$.

\section{Gaussian MSE Setting\label{sec:Gaussian-MSE-Setting}}

In the following theorem we show that, under MSE, a Gaussian source
belongs to $\mathcal{Z}_{ZB}$. 
\begin{thm}
\label{thm:General_CMS-1}(i) CMS achieves the \textbf{complete} RD
region for the symmetric 3-descriptions quadratic Gaussian setup shown
in Fig. \ref{fig:3_des_new}. 

(ii) The VKG encoding scheme cannot achieve all the points in the
region, i.e., \textup{$\overline{\mathcal{RD}}_{VKG}\subset\overline{\mathcal{RD}}_{CMS}$. }\end{thm}
\begin{rem}
It follows from (i) and (ii) that $\mathcal{RD}_{VKG}\subset\mathcal{RD}_{CMS}$
for the $L-$channel quadratic Gaussian MD problem $\forall L>2$. 
\end{rem}

\begin{proof}
Proof of (i) is straightforward and follows directly from the proof
of Theorem 1. Hence, we only prove (ii). Specifically, we show that,
a Gaussian random variable, under MSE, belongs to $\mathcal{Z}_{ZB}$.
(ii) then follows directly from Theorem 1. 

Consider the 2-description quadratic Gaussian problem. It follows
from Ozarow's results (see also \cite{EGC}) that, if $D_{12}\leq D_{1}+D_{2}-1$,
then the following rate region is achievable (and complete):
\begin{eqnarray}
R_{\mathcal{K}} & \geq & \frac{1}{2}\log\frac{1}{D_{\mathcal{K}}},\,\,\mathcal{K}\in\{1,2,12\}\label{eq:high_dist_region-1}
\end{eqnarray}
i.e., there is no excess rate incurred due to encoding the source
using two descriptions. Observe that the excess sum rate term in the
ZB must be set to zero to achieve the above rate-region. We will show
that, if we restrict the optimization to conditionally independent
joint densities, then it is impossible to simultaneously satisfy all
the distortions and achieve $I(U_{1};U_{2}|V_{12})=0$. Recall that
the ZB region achievable using any joint density $P(X,V_{12},U_{1},U_{2},U_{12})$
is given by:
\begin{eqnarray}
R_{i} & \geq & I(X;V_{12},U_{i})\,\, i\in\{1,2\}\nonumber \\
R_{1}+R_{2} & \geq & I(X;V_{12})+I(U_{1};U_{2}|V_{12})\nonumber \\
 &  & +I(X;V_{12},U_{1},U_{2},U_{12})\nonumber \\
D_{\mathcal{K}} & \geq & E\left[(X-\psi_{\mathcal{K}}(U_{\mathcal{K}}))\right],\,\,\mathcal{K}\in\{1,2,12\}\label{eq:}
\end{eqnarray}
Let us consider the corner point $P_{0}\triangleq(R_{1},R_{2})=(\frac{1}{2}\log\frac{1}{D_{1}},\frac{1}{2}\log\frac{D_{1}}{D_{12}})$
for some $(D_{1},D_{2},D_{12})$ satisfying $D_{12}\leq D_{1}+D_{2}-1$
and show that this point is not achievable by the ZB scheme when we
restrict the joint densities to satisfy (\ref{eq:ZZB}). First, as
$I(X;V_{12},U_{1},U_{2},U_{12})\geq\frac{1}{2}\log\frac{1}{D_{12}}$,
$P_{0}$ can be achieved only by joint densities that satisfy $I(X;V_{12})=0$.
Hence, to prove the theorem, it is sufficient to show that $P_{0}$
is not achievable when we restrict the optimization to joint densities
satisfying (\ref{eq:ZZB}) and $I(X;V_{12})=0$. 

Let $P(V_{12},U_{1},U_{2},U_{12},X)$ be any such joint density and
let $\mathcal{V}_{12}$ be the alphabet for $V_{12}$. Then the associated
ZB achievable region can be rewritten as:

\begin{eqnarray}
R_{i} & \geq & \sum_{v_{12}\in\mathcal{V}_{12}}P(v_{12})I(X;U_{i}|V_{12}=v_{12}),\,\, i\in\{1,2\}\nonumber \\
R_{1}+R_{2} & \geq & \sum_{v_{12}\in\mathcal{V}_{12}}P(v_{12})\Bigl[I(U_{1};U_{2}|V_{12}=v_{12})\nonumber \\
 &  & +I(X;U_{1},U_{2},U_{12}|V_{12}=v_{12})\Bigr]\nonumber \\
D_{\mathcal{K}} & \geq & E\left[(X-\psi_{\mathcal{K}}(U_{\mathcal{K}}))^{2}\right],\,\,\mathcal{K}\in\{1,2,12\}\nonumber \\
 & = & E\left[E\left[(X-\psi_{\mathcal{K}}(U_{\mathcal{K}}))^{2}\Bigl|V_{12}\right]\right]\label{eq:ZB_v12_split-1}
\end{eqnarray}
We will next show that the optimization can be further restricted
to joint densities $P(X,V_{12})Q(\tilde{U}_{1},\tilde{U}_{2},\tilde{U}_{12}|X,V_{12})$
such that $(X,\tilde{U}_{1},\tilde{U}_{2},\tilde{U}_{12})$ are jointly
Gaussian given $V_{12}=v_{12}$, $\forall v_{12}\in\mathcal{V}_{12}$
and satisfying $Q(\tilde{U}_{1},\tilde{U}_{2}|X,V_{12})=Q(\tilde{U}_{1}|X,V_{12})Q(\tilde{U}_{2}|X,V_{12})$.
First, note that, as $I(X;V_{12})=0$, $P(X|V_{12}=v_{12})$ is Gaussian
$\forall v_{12}\in\mathcal{V}_{12}$. Next, recall that $P_{0}$ is
obtained by first minimizing $R_{1}$ followed by minimizing $R_{2}$
given $R_{1}$ subject to all the distortion constraints. From (\ref{eq:ZB_v12_split-1}),
we have $R_{1}=\min\sum_{v_{12}\in\mathcal{V}_{12}}P(v_{12})I(X;U_{1}|V_{12}=v_{12})$,
where the minimization is over all joint densities $P(X,V_{12},U_{1})$
satisfying the distortion constraint for $D_{1}$. 

Let $P(X,V_{12},U_{1})$ be any joint density satisfying the distortion
constraint for $D_{1}$. Consider the joint density generated as $Q(X,V_{12},\tilde{U}_{1})=P(X,V_{12})Q(\tilde{U}_{1}|X,V_{12})$
where $(X,\tilde{U}_{1})$ are jointly Gaussian given $V_{12}=v_{12}$
and $K_{X,\tilde{U}_{1}|V_{12}=v_{12}}=K_{X,U_{1}|V_{12}=v_{12}}$,
$\forall v_{12}\in\mathcal{V}_{12}$. Observe that $Q(\cdot)$ also
satisfies the distortion constraint for $D_{1}$. As a Gaussian density
over the relevant random variables maximizes the conditional entropy
for a fixed covariance matrix \cite{Thomas}, it follows that $I(X;U_{1}|V_{12}=v_{12})\geq I(X;\tilde{U}_{1}|V_{12}=v_{12})$,
$\forall v_{12}\in\mathcal{V}_{12}$. Hence, to achieve minimum $R_{1}$,
it is sufficient to restrict the optimization to densities wherein
$(X,U_{1})$ are jointly Gaussian given $V_{12}$. 

Next consider minimizing $R_{2}$ given $R_{1}$. From (\ref{eq:ZB_v12_split-1}),
we have:
\begin{eqnarray}
R_{2} & = & \min\Bigl\{\sum_{v_{12}\in\mathcal{V}_{12}}P(v_{12})\Bigl[I(\tilde{U}_{1};U_{2}|V_{12}=v_{12})\nonumber \\
 &  & +I(X;\tilde{U}_{1},U_{2},U_{12}|V_{12}=v_{12})\Bigr]-R_{1}\Bigr\}
\end{eqnarray}
where the minimization is over all joint densities $P(X,V_{12},U_{1})P(U_{2},U_{12}|X,V_{12},U_{1})$
satisfying (\ref{eq:ZZB}) and $I(X;V_{12})=0$ and where $(X,U_{1})$
are jointly Gaussian given $V_{12}=v_{12}$, $\forall v_{12}\in\mathcal{V}_{12}$
(required to minimize $R_{1}$). It is easy to show using similar
arguments that the above minimization is again achieved by a joint
density where $(X,U_{1},U_{2},U_{12})$ are jointly Gaussian given
$V_{12}=v_{12}$ and such that $Q(U_{1},U_{2}|X,V_{12})=Q(U_{1}|X,V_{12})Q(U_{2}|X,V_{12})$
$\forall v_{12}\in\mathcal{V}_{12}$. Hence, to achieve $P_{0}$ using
a joint density that satisfies (\ref{eq:ZZB}), it is sufficient to
optimize the rates over joint densities satisfying the following properties:

1) $(X,U_{1},U_{2},U_{12})$ are jointly Gaussian given $V_{12}=v_{12}$
$\forall v_{12}\in\mathcal{V}_{12}$

2) $I(X;V_{12})=0$

3) $I(U_{1};U_{2}|X,V_{12})=0$

4) $P(X,V_{12},U_{1},U_{2},U_{12})$ satisfies all the distortion
constraints

Observe that, for any joint density that satisfies all the above properties,
it is impossible to achieve $I(U_{1};U_{2}|V_{12})=0$. Therefore,
the excess sum rate term in the ZB scheme is non-zero, concluding
that $P_{0}$ is not achievable by the ZB scheme using any independent
quantization mechanism. Therefore, a Gaussian random variable under
MSE belongs to $\mathcal{Z}_{ZB}$, proving the theorem.
\end{proof}
Note that, as $\mathcal{Z}_{ZB}\subseteq\mathcal{Z}_{EC}$, a Gaussian
source under MSE belongs to $\mathcal{Z}_{EC}$. Hence, the `correlated
quantization' scheme (an extreme special case of VKG) which has been
proven to be complete for several cross-sections of the $L-$descriptions
quadratic Gaussian MD problem \cite{Jun_Chen_ind_central}, is strictly
suboptimal in general.

\section{Points on the boundary\label{sub:Points-on-the}}

Before stating the results formally, we review Ozarow's result for
the 2-descriptions MD setting. Ozarow showed that the complete region
for the 2-descriptions Gaussian MD problem can be achieved using a
`correlated quantization' scheme which imposes the following joint
distribution for $(U_{1},U_{2},U_{12})$ in the EC scheme:
\begin{eqnarray}
U_{1}=X+W_{1}\nonumber \\
U_{2}=X+W_{2}\label{eq:Ozarow_Result-1}
\end{eqnarray}
$U_{12}=E(X|U_{1},U_{2})$, where $W_{1}$ and $W_{2}$ are zero mean
Gaussian random variables independent of $X$ with covariance matrix
$K_{W_{1}W_{2}}$, and the functions $\psi_{\mathcal{K}}(U{}_{\mathcal{K}})$
are given by the respective MSE optimal estimators, eg., $\psi_{1}(U_{1})=E\left[X|U_{1}\right]$.
The covariance matrix $K_{W_{1}W_{2}}$ is set to satisfy all the
distortion constraints. Specifically, the optimum $K_{W_{1}W_{2}}$
is given by:
\begin{equation}
K_{W_{1}W_{2}}=\left[\begin{array}{cc}
\sigma_{1}^{2} & \rho_{12}\sigma_{1}\sigma_{2}\\
\rho_{12}\sigma_{1}\sigma_{2} & \sigma_{2}^{2}
\end{array}\right]\label{eq:Ozarow_K-1}
\end{equation}
where $\sigma_{i}^{2}=\frac{D_{i}}{1-D_{i}}\,\, i\in\{1,2\}$ and
the optimum $\rho_{12}$, denoted by $\rho_{12}^{*}$, is given by
(see \cite{Zamir}):
\begin{eqnarray}
\rho_{12}^{*} & = & \begin{cases}
-\frac{\sqrt{\pi D_{12}^{2}+\gamma}-\sqrt{\pi D_{12}^{2}}}{(1-D_{12})\sqrt{D_{1}D_{2}}} & D_{12}\leq D_{12}^{max}\\
0 & D_{12}\geq D_{12}^{max}
\end{cases}\nonumber \\
\gamma & = & (1-D_{12})\Bigl[(D_{1}-D_{12})(D_{2}-D_{12})\nonumber \\
 &  & +D_{12}D_{1}D_{2}-D_{12}^{2}\Bigr]\nonumber \\
D_{12}^{max} & = & D_{1}D_{2}/(D_{1}+D_{2}-D_{1}D_{2})\nonumber \\
\pi & = & (1-D_{1})(1-D_{2})\label{eq:other_defn-1}
\end{eqnarray}
We denote the complete Gaussian-MSE $L$-descriptions region by $\mathcal{RD}_{G}^{L}$.
The characterization of $\mathcal{RD}_{G}^{2}$ is given in \cite{EGC}
(see also \cite{Zamir}) and we omit restating it explicitly here
for brevity. 

In this section we show that CMS achieves the complete RD region for
several cross-sections of the general quadratic Gaussian $L-$channel
MD problem. We again begin with the 3-descriptions case and then extend
the results to the $L$ channel framework. Recall the setup shown
in Fig. \ref{fig:3_des_new}, i.e, a cross-section of the general
3-descriptions rate-distortion region wherein we impose constraints
only on distortions $(D_{1},D_{2},D_{3},D_{12},D_{23})$ and set the
rest of the distortions, $(D_{13},D_{123})$ to $1$. Here we consider
the general asymmetric case, i.e. $D_{1}\neq D_{3}$ and $D_{12}\neq D_{23}$
and show that the CMS scheme achieves the complete rate region in
several distortion regimes. 

In the following theorem, without loss of generality we assume that
$D_{1}\leq D_{3}$. If $D_{3}\leq D_{1}$, then the theorem holds
by interchanging `$1$' and `$3$' everywhere. Let $D_{12}$ be any
distortion such that $D_{12}\leq\min\{D_{1},D_{2}\}$. We define $D_{23}^{*}=D_{23}^{*}(D_{1},D_{2},D_{3},D_{12})$
as:
\begin{equation}
D_{23}^{*}=\frac{\sigma_{2}^{2}\sigma_{3}^{2}\left(1-\rho{}^{2}\right)}{\sigma_{2}^{2}\sigma_{3}^{2}\left(1-\rho{}^{2}\right)+\sigma_{2}^{2}+\sigma_{3}^{2}-2\sigma_{2}\sigma_{3}\rho}\label{eq:defn_D23-1}
\end{equation}
where $\sigma_{i}^{2}=\frac{D_{i}}{1-D_{i}}$ $i\in\{2,3\}$ and
\begin{equation}
\rho=\rho_{12}^{*}\frac{\sigma_{1}}{\sigma_{3}}\label{eq:defn_rho}
\end{equation}
where $\rho_{12}^{*}$ is defined in (\ref{eq:other_defn-1}). In
the following theorem, we will show that CMS achieves the complete
rate-region if $D_{23}=D_{23}^{*}$.
\begin{thm}
\label{thm:Sum_Rate_Gauss}For the setup shown in Fig. \ref{fig:3_des_new},
let $D_{1}\leq D_{3}$. Then,

(i) CMS achieves the complete rate-region if:
\begin{eqnarray}
D_{23} & = & D_{23}^{*}(D_{1},D_{2},D_{3},D_{12})\label{eq:sum_rate_thm1}
\end{eqnarray}
where $D_{23}^{*}$ is defined in (\ref{eq:defn_D23-1}). The rate
region is given by:
\begin{eqnarray}
R_{i} & \geq & \frac{1}{2}\log\frac{1}{D_{i}}\,\, i\in\{1,2,3\}\nonumber \\
R_{1}+R_{2} & \geq & \frac{1}{2}\log\frac{1}{D_{1}D_{2}}+\delta(D_{1},D_{2},D_{12})\nonumber \\
R_{2}+R_{3} & \geq & \frac{1}{2}\log\frac{1}{D_{2}D_{3}}+\delta(D_{2},D_{3},D_{23})\label{eq:sum_rate_thm2}
\end{eqnarray}
where $\delta(\cdot)$ is defined by:
\begin{equation}
\delta(D_{1},D_{2},D_{12})=\frac{1}{2}\log\left(\frac{1}{1-(\rho_{12}^{*})^{2}}\right)\label{eq:delta_defn}
\end{equation}
where $\rho_{12}^{*}$ is defined in (\ref{eq:other_defn-1}).

(ii) Moreover, CMS achieves the minimum sum-rate if one of the following
hold:

(a) For a fixed $D_{12}$, $D_{23}\geq D_{23}^{*}(D_{1},D_{2},D_{3},D_{12})$

(b) For a fixed $D_{23}$, $D_{12}\in\{D_{12}:\delta(D_{2},D_{3},D_{23})\geq\delta(D_{1},D_{2},D_{12})\}$\end{thm}
\begin{rem}
We note that the above rate region \textit{cannot} be achieved by
VKG. We omit the details of the proof here as it can be proved in
same lines as the proof of Theorem \ref{thm:General_CMS-1}.
\end{rem}

\begin{rem}
An achievable rate-distortion region can be derived for general distortions
using the encoding principles we derive as part of this proof. However,
it is hard to prove outer bounds if the conditions in (\ref{eq:sum_rate_thm1})
are not satisfied and hence we omit stating the results explicitly
here. 
\end{rem}

\begin{rem}
Both the CMS and the VKG schemes achieve the complete rate region
when $D_{12}\geq D_{12}^{max}$ and $D_{23}\geq D_{23}^{max}$, where
$D_{12}^{max}$ and $D_{23}^{max}$ are defined in (\ref{eq:other_defn-1}).
It is easy to show that in this case an independent quantization scheme
is optimal and the complete achievable rate-region is given by $R_{i}\geq\frac{1}{2}\log\frac{1}{D_{i}}\,\, i\in\{1,2,3\}$. 
\end{rem}

\begin{rem}
It is easy to verify that CMS achieves the minimum sum-rate whenever
$D_{12}=D_{23}$ for any $D_{1},D_{3}$. \end{rem}
\begin{proof}
We begin with the proof of (i). The proof of (ii) then follows almost
directly from (i). First we show the converse, which is quite obvious.
Conditions on $R_{i}$ follow from the converse to the source coding
theorem. Conditions on $R_{1}+R_{2}$ and $R_{2}+R_{3}$ follow from
Ozarow's result, to achieve $(D_{1},D_{2},D_{12})$ using descriptions
$\{1,2\}$ and to achieve $(D_{2},D_{3},D_{23})$ using descriptions
$\{2,3\}$ at the respective decoders.

We next prove that CMS achieves the rate region in (\ref{eq:sum_rate_thm2})
if (\ref{eq:sum_rate_thm1}) holds. We first give an intuitive argument
to explain the encoding scheme. Description 3 carries an RD-optimal
quantized version of $X$ (which achieves distortion $D_{3}$). Description
1 carries all the bits embedded in description 3 along with `refinement
bits' which assist in achieving distortion $D_{1}\leq D_{3}$. This
entails no loss in optimality as a Gaussian source is successively
refinable under MSE \cite{Successive_Refinement}. Description 2 then
carries a quantized version of the source which is correlated with
the information in descriptions 1 and 3. We will show that if $D_{23}=D_{23}^{*}(D_{1},D_{2},D_{3},D_{12})$,
then the correlations can be set such that description 2 is optimal
with respect to both descriptions 1 and 3. 

Formally, to achieve the rate region in (\ref{eq:sum_rate_thm2}),
we set the auxiliary random variables in the CMS coding scheme as
follows:
\begin{eqnarray}
V_{13} & = & X+W_{1}+W_{3}\nonumber \\
U_{3} & = & V_{13}\nonumber \\
U_{1} & = & X+W_{1}\nonumber \\
U_{2} & = & X+W_{2}\nonumber \\
U_{12}=\Phi &  & U_{23}=\Phi\label{eq:Sum_rate_pf_3}
\end{eqnarray}
and the functions $\psi(\cdot)$ as the respective MSE optimal estimators,
where $W_{1},W_{2},W_{3}$ are zero mean Gaussian random variables
independent of $X$ with a covariance matrix:
\begin{equation}
K_{W_{1}W_{2}W_{3}}=\left[\begin{array}{ccc}
\tilde{\sigma}_{1}^{2} & \rho_{12}\tilde{\sigma}_{1}\tilde{\sigma}_{2} & 0\\
\rho_{12}\tilde{\sigma}_{1}\tilde{\sigma}_{2} & \tilde{\sigma}_{2}^{2} & 0\\
0 & 0 & \tilde{\sigma}_{3}^{2}
\end{array}\right]\label{eq:Sum_rate_pf_4}
\end{equation}
where $\tilde{\sigma}_{1}^{2}=\sigma_{1}^{2}=\frac{D_{1}}{1-D_{1}}$,
$\tilde{\sigma}_{2}^{2}=\sigma_{2}^{2}=\frac{D_{2}}{1-D_{2}}$, $\tilde{\sigma}_{3}^{2}=\sigma_{3}^{2}-\sigma_{1}^{2}=\frac{D_{3}}{1-D_{3}}-\frac{D_{1}}{1-D_{1}}$.
The correlation coefficient $\rho_{12}$ is set to achieve distortion
$D_{12}$, i.e. $\rho_{12}=\rho_{12}^{*}$ defined in (\ref{eq:other_defn-1}).
Let us denote by $W_{13}=W_{1}+W_{3}$. Observe that the encoding
for descriptions 2 and 3 resembles Ozarow's correlated quantization
scheme with $U_{2}=X+W_{2}$ and $U_{3}=X+W_{13}$. Let us denote
the correlation coefficient between $W_{2}$ and $W_{13}$ be $\rho$.
We have the following equation relating $\rho_{12}$ and $\rho$ (which
is equivalent to (\ref{eq:defn_rho})):
\begin{equation}
\rho_{12}^{*}\tilde{\sigma}_{1}=\rho\sqrt{\tilde{\sigma}_{1}^{2}+\tilde{\sigma}_{3}^{2}}\label{eq:rho_defn2}
\end{equation}
Note that the above relation is derived using the independence of
$W_{2}$ and $(W_{1},W_{3})$, which follows from our choice of $K_{W_{1}W_{2}W_{3}}$.
Hence the minimum distortion $D_{23}$ achievable using the above
choice for the joint density of the auxiliary random variables is
given by:
\begin{eqnarray}
D_{23} & = & \mbox{Var}(X|U_{2},U_{3},V_{13})\nonumber \\
 & = & \mbox{Var}(X|U_{2},V_{13})\nonumber \\
 & = & D_{23}^{*}\label{eq:min_D23}
\end{eqnarray}

We next derive the rates required by this choice of $K_{W_{1}W_{2}W_{3}}$.
Direct application of Theorem \ref{thm:main} using the above joint
density leads to the following achievable rate region for any given
distortions $D_{1},D_{2},D_{3},D_{12},D_{23}$:
\begin{eqnarray*}
R_{13}^{''} & \geq & \frac{1}{2}\log\frac{1}{D_{3}}\\
R_{2}^{'} & \geq & \frac{1}{2}\log\frac{1}{D_{2}}\\
R_{1}^{'}+R_{13}^{''} & \geq & \frac{1}{2}\log\frac{1}{D_{1}}\\
R_{2}^{'}+R_{13}^{''} & \geq & H(V_{13})+H(U_{2})-H(V_{13},U_{2}|X)\\
 & = & H(U_{3})+H(U_{2})-H(U_{3},U_{2}|X)\\
 & = & \frac{1}{2}\log\frac{1}{D_{3}D_{2}}+\frac{1}{2}\log\left(\frac{1}{1-\rho{}^{2}}\right)\\
 & = & \frac{1}{2}\log\frac{1}{D_{3}D_{2}}+\delta(D_{2},D_{3},D_{23}^{*})
\end{eqnarray*}
\begin{eqnarray*}
R_{1}^{'}+R_{2}^{'}+R_{13}^{''} & \geq & H(V_{13})+H(U_{1}|V_{13})+H(U_{2})\\
 &  & -H(U_{1},V_{13},U_{2}|X)\\
 & = & I(X;U_{1},V_{13})+I(U_{2};X,U_{1},V_{13})\\
 & =^{(a)} & I(X;U_{1})+I(X;U_{2})\\
 &  & +I(U_{2};U_{1},V_{13}|X)\\
 & = & I(X;U_{1})+I(X;U_{2})\\
 &  & +I(U_{2};U_{1},U_{3}|X)\\
 & =^{(b)} & I(X;U_{1})+I(X;U_{2})\\
 &  & +I(W_{2};W_{1},W_{1}+W_{3})\\
 & = & I(X;U_{1})+I(X;U_{2})+I(W_{2};W_{1})\\
 &  & +I(W_{2};W_{3}|W_{1})\\
 & =^{(c)} & I(X;U_{1})+I(X;U_{2})+I(W_{2};W_{1})\\
 & = & \frac{1}{2}\log\frac{1}{D_{1}D_{2}}+\frac{1}{2}\log\left(\frac{1}{1-(\rho_{12}^{*})^{2}}\right)\\
 & = & \frac{1}{2}\log\frac{1}{D_{1}D_{2}}+\delta(D_{1},D_{2},D_{12})
\end{eqnarray*}
\begin{eqnarray}
R_{1} & = & R_{13}^{''}+R_{1}^{'}\nonumber \\
R_{2} & = & R_{2}^{'}\nonumber \\
R_{3} & = & R_{13}^{''}\label{eq:CMS_Gauss_ach_region}
\end{eqnarray}
where $(a)$ follows from the Markov chain $X\leftrightarrow U_{1}\leftrightarrow V_{13}$,
$(b)$ from the independence of $X$ and $(W_{1},W_{2},W_{3})$ and
$(c)$ from the independence of $W_{3}$ and $(W_{1},W_{2})$. 

At a first glance, it might be tempting to conclude that the region
for the tuple $(R_{1},R_{2},R_{3})$ in (\ref{eq:CMS_Gauss_ach_region})
is equivalent to the region given by (\ref{eq:sum_rate_thm2}). This
is not the case in general as the equations in (\ref{eq:CMS_Gauss_ach_region})
have an implicit constraint on the auxiliary rates $R_{13}^{''},R_{1}^{'},R_{2}^{'}\geq0$.
However, we will show that if $D_{3}\geq D_{1}$, then the two regions
are indeed equivalent. We denote the rate region given in (\ref{eq:sum_rate_thm2})
by $\mathcal{R}$ and the region in (\ref{eq:CMS_Gauss_ach_region})
by $\mathcal{R}^{*}$. Clearly, $\mathcal{R}^{*}\subseteq\mathcal{R}$,
as any $(R_{1},R_{2},R_{3})$ that satisfies (\ref{eq:CMS_Gauss_ach_region})
also satisfies (\ref{eq:sum_rate_thm2}). We need to show that $\mathcal{R}^{*}\supseteq\mathcal{R}$.
Towards proving this claim, note that both $\mathcal{R}$ and $\mathcal{R}^{*}$
are convex regions bounded by hyper-planes. Hence, it is sufficient
for us to show that all the corner points of $\mathcal{R}$ lie in
$\mathcal{R}^{*}$. Clearly, $\mathcal{R}$ has 6 corner points denoted
by $P_{ijk}$ $i,j,k\in\{1,2,3\}$ defined as:
\begin{eqnarray}
P_{ijk} & = & \{r_{i},r_{j},r_{k}\}\nonumber \\
r_{i} & = & \min R_{i}\nonumber \\
r_{j} & = & \min_{R_{i}=r_{i}}R_{j}\nonumber \\
r_{k} & = & \min_{R_{i}=r_{i},R_{j}=r_{j}}R_{k}
\end{eqnarray}
To prove $\mathcal{R}^{*}\supseteq\mathcal{R}$, we need to prove
that every corner point $(r_{1},r_{2},r_{3})\in\mathcal{R}$ is achieved
by some non-negative $(R_{13}^{''},R_{1}^{'},R_{2}^{'},R_{1},R_{2},R_{3})\in\mathcal{R}^{*}$
such that $R_{i}=r_{i},\,\, i\in\{1,2,3\}$. We set $R_{13}^{''}=R_{3}=r_{3}$
and $R_{2}^{'}=R_{2}=r_{2}$ and show that we can always find $R_{1}^{'}\geq0$
satisfying (\ref{eq:CMS_Gauss_ach_region}) such that $R_{1}=R_{1}^{'}+R_{13}^{'}=r_{1}$.
Let us first consider the points $P_{213}=P_{231}$ given by:
\begin{eqnarray}
r_{1} & = & \frac{1}{2}\log\frac{1}{D_{1}}+\delta(D_{1},D_{2},D_{12})\nonumber \\
r_{2} & = & \frac{1}{2}\log\frac{1}{D_{2}}\nonumber \\
r_{3} & = & \frac{1}{2}\log\frac{1}{D_{3}}+\delta(D_{2},D_{3},D_{23})
\end{eqnarray}
This can be achieved by using the following auxiliary rates, $R_{2}^{'}=r_{2}$,
$R_{13}^{''}=r_{3}$ and 
\begin{eqnarray}
R_{1}^{'} & = & \frac{1}{2}\log\frac{D_{3}}{D_{1}}+\delta(D_{1},D_{2},D_{12})\nonumber \\
 &  & -\delta(D_{2},D_{3},D_{23})\nonumber \\
 & = & \frac{1}{2}\log\frac{(1-D_{1})D_{3}-(\rho_{12}^{*})^{2}D_{1}(1-D_{3})}{(1-D_{1})D_{1}(1-(\rho_{12}^{*})^{2})}
\end{eqnarray}
It is easy to verity that $R_{1}^{'}\geq0$ if $D_{3}\geq D_{1}$.
Hence $P_{213}=P_{231}\in\mathcal{R}^{*}$. Let us next consider the
points $P_{132}=P_{312}$ given by: 
\begin{eqnarray}
r_{1} & = & \frac{1}{2}\log\frac{1}{D_{1}}\nonumber \\
r_{2} & = & \frac{1}{2}\log\frac{1}{D_{2}}\nonumber \\
 &  & +\max\{\delta(D_{1},D_{2},D_{12}),\delta(D_{2},D_{3},D_{23})\}\nonumber \\
r_{3} & = & \frac{1}{2}\log\frac{1}{D_{3}}
\end{eqnarray}
Again it is easy to show that $(R_{13}^{''},R_{1}^{'},R_{2}^{'})=(r_{3},r_{1}-r_{3},r_{2})$
belongs to $\mathcal{R}^{*}$. Finally, we consider the remaining
two points $P_{123}$ and $P_{321}$. $P_{123}$ is given by:
\begin{eqnarray}
r_{1} & = & \frac{1}{2}\log\frac{1}{D_{1}}\nonumber \\
r_{2} & = & \frac{1}{2}\log\frac{1}{D_{2}}+\delta(D_{1},D_{2},D_{12})\nonumber \\
r_{3} & = & \frac{1}{2}\log\frac{1}{D_{3}}\nonumber \\
 &  & +\left(\delta(D_{2},D_{3},D_{23})-\delta(D_{1},D_{2},D_{12})\right)^{+}
\end{eqnarray}
where $x^{+}=\max\{x,0\}$. Consider the following auxiliary rates:
$R_{13}^{''}=r_{3}$, $R_{2}^{'}=r_{2}$ and $R_{1}^{'}=\frac{1}{2}\log\frac{D_{3}}{D_{1}}$.
Clearly the first three constraints in (\ref{eq:CMS_Gauss_ach_region})
are satisfied by these auxiliary rates. The following inequalities
prove that the last two constraints are also satisfied by these rates
and hence $P_{123}\in\mathcal{R}^{*}$.
\begin{eqnarray*}
R_{2}^{'}+R_{13}^{''} & = & r_{2}+r_{3}\\
 & = & \frac{1}{2}\log\frac{1}{D_{2}D_{3}}+\delta(D_{1},D_{2},D_{12})\\
 &  & +\left(\delta(D_{2},D_{3},D_{23})-\delta(D_{1},D_{2},D_{12})\right)^{+}\\
 & \geq & \frac{1}{2}\log\frac{1}{D_{2}D_{3}}+\delta(D_{2},D_{3},D_{23})
\end{eqnarray*}
\begin{eqnarray}
R_{2}^{'}+R_{1}^{'}+R_{13}^{''} & = & \frac{1}{2}\log\frac{1}{D_{1}D_{2}}+\delta(D_{1},D_{2},D_{12})\nonumber \\
 &  & +\left(\delta(D_{2},D_{3},D_{23})-\delta(D_{1},D_{2},D_{12})\right)^{+}\nonumber \\
 & \geq & \frac{1}{2}\log\frac{1}{D_{1}D_{2}}+\delta(D_{1},D_{2},D_{12})
\end{eqnarray}
Next consider $P_{321}$:
\begin{eqnarray}
r_{1} & = & \frac{1}{2}\log\frac{1}{D_{1}}\nonumber \\
 &  & +\left(\delta(D_{1},D_{2},D_{12})-\delta(D_{2},D_{3},D_{23})\right)^{+}\nonumber \\
r_{2} & = & \frac{1}{2}\log\frac{1}{D_{2}}+\delta(D_{2},D_{3},D_{23})\nonumber \\
r_{3} & = & \frac{1}{2}\log\frac{1}{D_{3}}
\end{eqnarray}
Using same arguments as before, it is easy to show that $P_{321}\in\mathcal{R}^{*}$
by using the following auxiliary rates: $R_{13}^{''}=r_{3}$, $R_{2}^{'}=r_{2}$
and $R_{1}^{'}=\frac{1}{2}\log\frac{D_{3}}{D_{1}}+\left(\delta(D_{1},D_{2},D_{12})-\delta(D_{2},D_{3},D_{23})\right)^{+}$.
Therefore, it follows that $\mathcal{R}=\mathcal{R}^{*}$ and hence
CMS achieves the complete rate region, proving (i). 

We next prove (ii)(a). It follows from (i) that the following rate
point is achievable $\forall D_{23}\geq D_{23}^{*}$:
\begin{eqnarray}
\left\{ R_{1},R_{2},R_{3}\right\}  & = & \Bigl\{\frac{1}{2}\log\frac{1}{D_{1}},\frac{1}{2}\log\frac{1}{D_{2}}+\delta(D_{1},D_{2},D_{12}),\nonumber \\
 &  & \frac{1}{2}\log\frac{1}{D_{3}}\Bigr\}
\end{eqnarray}
Also observe that $\forall D_{23}\geq D_{23}^{*}$, $\delta(D_{1},D_{2},D_{12})\geq\delta(D_{2},D_{3},D_{23})$
and hence a lower bound to the sum rate is $\frac{1}{D_{1}D_{2}D_{3}}+\delta(D_{1},D_{2},D_{12})$.
Therefore the above point achieves the minimum sum rate $\forall D_{23}\geq D_{23}^{*}$. 

The proof of (ii)(b) follows similarly by noting that if $D_{12}\in\{D_{12}:\delta(D_{2},D_{3},D_{23})\geq\delta(D_{1},D_{2},D_{12})\}$,
the minimum sum rate is given by $\frac{1}{D_{1}D_{2}D_{3}}+\delta(D_{2},D_{3},D_{23})$
which is achieved by the point:
\begin{eqnarray}
\left\{ R_{1},R_{2},R_{3}\right\}  & = & \Bigl\{\frac{1}{2}\log\frac{1}{D_{1}},\frac{1}{2}\log\frac{1}{D_{2}}+\delta(D_{2},D_{3},D_{12}),\nonumber \\
 &  & \frac{1}{2}\log\frac{1}{D_{3}}\Bigr\}
\end{eqnarray}
This proves the theorem. 
\end{proof}
It is interesting to observe that the optimum encoding scheme introduces
common codewords (creates an interaction) between descriptions 1 and
3, even though these two descriptions are never received simultaneously
at the decoder. While common codewords typically imply redundancy
in the system, in this case, introducing them allows for better co-ordination
between the descriptions leading to a smaller rate for the common
branch. Similar principles can be used to show that CMS achieves the
complete RD region for the $L-$channel quadratic Gaussian MD problem,
for several distortion regimes. 

\begin{figure}
\centering\includegraphics[scale=0.25]{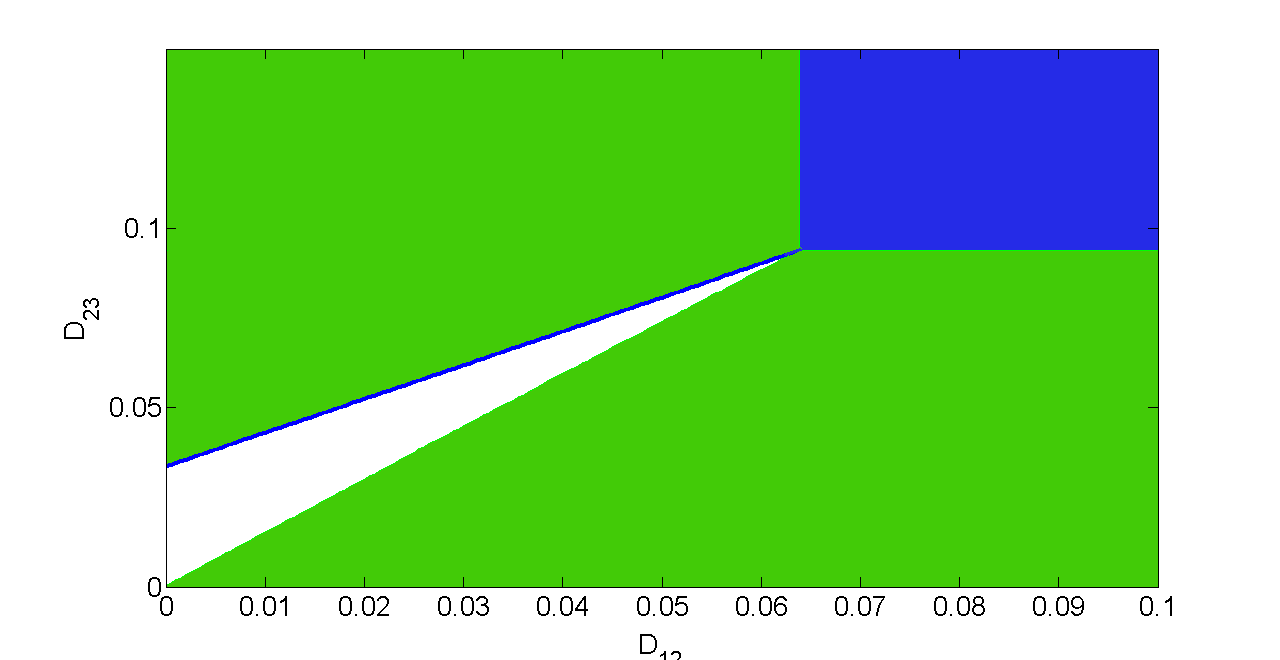}

\caption{Example: This figure denotes the regime of distortions wherein the
CMS scheme achieves the complete rate region and the minimum sum rate.
Here $D_{1}=0.1,D_{2}=0.15$ and $D_{3}=0.2$. The blue points correspond
to the region of distortions wherein the CMS scheme achieves the complete
rate-region and the green points represent the region where the CMS
scheme achieves the minimum sum rate \label{fig:Example:-This-figure}.}
\end{figure}

\begin{example}
We consider an asymmetric setting where $D_{1}=0.1,D_{2}=0.15$ and
$D_{3}=0.2$. Fig. \ref{fig:Example:-This-figure} shows the regime
of distortions where CMS achieves the complete rate-region and minimum
sum rate. The blue region corresponds to the set of distortion pairs
$(D_{12},D_{23})$ wherein the CMS rate-region is complete. The green
region denotes the minimum sum rate points. It is clearly evident
from the figure that CMS achieves the minimum sum rate for a fairly
large regime of distortions. 
\end{example}

\section{Conclusion}

In this paper, we showed that CMS achieves a strictly larger region
compared to VKG for a general class of sources and distortion measures,
which includes the quintessential setting of Gaussian source under
mean squared error. As a consequence, it follows that the `correlated
quantization' scheme (an extreme special case of VKG), is strictly
suboptimal in general. We also showed that CMS achieves the complete
rate region for the 3-description symmetric cross-section and several
asymmetric cross-sections of the setup shown in Fig. \ref{fig:3_des_new}.

\bibliographystyle{unsrt}
\bibliography{Journal_Bibtex}

\end{document}